\documentclass[11pt]{article}
\usepackage{epsfig}
\usepackage{amsmath,amssymb,amsthm}
\usepackage{graphicx}
\usepackage{flafter}
\usepackage{parskip}
\usepackage{authblk}

\usepackage[round]{natbib}
\usepackage{color,hyperref,xcolor}
\hypersetup{colorlinks=true,urlcolor=blue, citecolor=purple}
\usepackage{cleveref}

\usepackage[margin=1cm,labelfont=bf]{caption}

\newtheoremstyle{general}
{3mm} 
{3mm} 
{\it} 
{} 
{\bfseries} 
{.} 
{.5em} 
{} 

\def\I{\mathbb{I}}

\newcommand{\abs}[1]{\left\lvert #1 \right\rvert}
\newcommand{\norm}[1]{\left\| #1 \right\|}
\newcommand{\cov}{\mathrm{Cov}}

\newcommand{\dash}{^{\prime}}

\newcommand{\unit}{\ \mu gm^{-3}}
\newcommand{\ppm}{\mathrm{PM}_{2.5}}

\theoremstyle{general}

\newtheorem{theorem}{Theorem}

\numberwithin{equation}{section}
\numberwithin{theorem}{section}
\numberwithin{lemma}{section}
\numberwithin{defn}{section}
\numberwithin{corollary}{section}

\usepackage{lipsum}
\usepackage{fancyhdr}
\fancypagestyle{title}{%
  \setlength{\headheight}{22pt}%
  \fancyhf{}
  \fancyfoot[C]{\thepage}
  \fancyhead[L]{}
}%

\begin{document}

\pagenumbering{arabic}

\title{Spatio-temporal models with space-time interaction and their applications to air pollution data}
\author{Soudeep Deb and Ruey S Tsay \\ University of Chicago}

\maketitle
\thispagestyle{title}

\begin{abstract}
\noindent
It is of utmost importance to have a clear understanding of the status of air pollution and to provide forecasts and insights about the air quality to the general public and researchers in environmental studies. Previous studies of spatio-temporal models showed that even a short-term exposure to high concentrations of atmospheric fine particulate matters can be hazardous to the health of ordinary people. In this study, we develop a spatio-temporal model with space-time interaction for air pollution data ($\ppm$). The proposed model uses a parametric space-time interaction component along with the spatial and temporal components in the mean structure, and introduces a random-effects component specified in the form of zero-mean spatio-temporal processes. For application, we analyze the air pollution data ($\ppm$) from 66 monitoring stations across Taiwan. 

\vspace{9pt}
\end{abstract}

\def\thefigure{\arabic{figure}}
\def\thetable{\arabic{table}}

\renewcommand{\theequation}{\thesection.\arabic{equation}}

\fontsize{12}{14pt plus.8pt minus .6pt}\selectfont

\setcounter{section}{0} 
\setcounter{equation}{0} 

\newpage

\section{Introduction}
\label{sec:introduction}

The effect of air pollution on public health, vegetation, and more generally, on the human society and the ecosystem has been a burning issue in recent years. Several epidemiological studies have established that the particulate matters (PM) are linked to a range of serious cardiovascular, respiratory, and visibility problems. Detailed discussions can be found in \citet{pope95}, \citet{jerrett13}, \citet{blangiardo16},  and \citet{thurston16}. Taking the severe effect of the PM into account,  the Environmental Protection Agency (EPA) of the United States of America (USA) provided in 1997 new regulations that established National Ambient Air Quality Standards (NAAQS) for PM with aerodynamic diameters less than $2.5$ micron. This is usually measured in units of micrograms per cubic meter ($\unit$), and we will denote it as $\ppm$ henceforth. According to NAAQS, the hourly average of the $\ppm$ concentration should not be over $35\ \unit$. However, the real situation often goes beyond the standard. A quick example is the data analyzed in this study. They are obtained from 66 different monitoring stations in Taiwan, for a span of 10 years (from 2006 to 2015) and the median of the $\ppm$ values is slightly above $37\ \unit$. The reader is also referred to the earlier study by \citet{mayer99} who discussed how the air quality is deteriorating in different cities across the world. All in all, there is a growing demand to identify the main factors that contribute to the air pollution.


A brief discussion on PM is in order. 
Generally speaking, $\ppm$ contains different particles either emitted directly or formed in the atmosphere from gaseous emissions. The examples include sulfates formed from sulfur dioxide ($SO_2$) emissions, nitrates formed from $NOx$ emissions, and carbon formed from organic gas emissions. Now, the rates of conversion of gases to particles are often reliant on different regional and temporal factors, including the topography, the land cover and several seasonal climatic variables, and as a consequence, the $\ppm$ concentrations are also affected by these variables. This immediately increases the need for a spatio-temporal model to assess the air quality. A good model can provide better predictions and that would in turn help in determining an efficient strategy to control the air pollution.

The spatial and spatio-temporal modeling of air pollutants started early in the past century. \citet{elsom78} studied the spatial correlation fields for air pollution in an urban area while different geostatistical space-time models have been applied to examine the trend in deposition of atmospheric pollutants in \citet{eynon83}, \citet{bilonick85}, \citet{rouhani89}, and \citet{vyas97}. Other earlier notable works in this regard include  \citet{guttorp94}, \citet{haas95}, and \citet{carroll97}. Most of these approaches rely on a spatio-temporal random field where the spatial or temporal dependences are incorporated in either the mean function or in the error process, and the parameters are estimated using frequentist procedures. A nice discussion on the geostatistical space-time models can be found in \citet{kyriakidis99}.

In comparison, many 21st century studies in the related problem have made use of hierarchical Bayesian approaches for spatial prediction of air pollution. For example, \citet{sun00} analyzed the $\mathrm{PM}_{10}$ (particulate matters with diameter less than $10\ \unit$) concentrations in Vancouver and developed posterior predictive distributions using Bayesian techniques. \citet{kibria02}, on the other hand, used a multivariate setup to analyze $\ppm$ concentrations in Philadelphia and developed spatial prediction methodology in a Bayesian context for this purpose. For a related problem, in order to predict $\mathrm{PM}_{10}$ concentrations in London, \citet{shaddick02} proposed a short-term space-time modeling technique. 

In a hierarchical Bayesian setting, \citet{sahu05} modeled the spatial structure using principal kriging functions and the time component by a random walk process to present a short-term forecasting analysis of $\ppm$ data in New York City. There are two other notable works by the same authors. \citet{sahu06} used indicators for urban or rural sites to employ different spatio-temporal processes in the error structure for modeling the $\ppm$ series of several states in the Midwest of the 
United States (US). 
In a later paper, \citet{sahu09} developed a space-time model, that includes a spatially varying regression term along with an auto-regressive term in the mean structure, to analyze the ozone concentrations in the eastern states of the US. \citet{berrocal10} extended one of their earlier works to introduce a bivariate downscaler and provided a flexible class of space-time assimilation models. On the other hand, \citet{cameletti11} provided a nice discussion on the comparison of available space-time models using data from Piemonte, Italy. 

In this study, we consider the problem of developing a new spatio-temporal model, with the main focus on identifying if there is any space-time interaction in the behavior of the $\ppm$ concentrations. An interaction means that the temporal trend of the pollution is more similar for sites closer in a spatial scale. An early study in this regard is by \citet{wikle98}. The authors developed a hierarchical Bayesian model that allowed an interaction in space and time. However, except for that and a few related works, there have been very few efforts to quantify the space-time interaction for air pollution data, albeit there is a vast literature on the spatio-temporal modeling for the same, as we have already discussed.

Note that the problem of identifying the space-time interaction is not at all specific to the air pollution data. Rather, it has been studied in several other fields, ranging from seismology, epidemiology, criminology to transportation research. \citet{meyer16} provided a nice discussion on the tests for space-time interaction in problems related to medical studies. The most popular techniques in this regard are Knox test, Mantel test, and space-time $K$-function analysis. These tests mainly revolve around test statistics of the form $T=\sum_{j\ne i}a_{ij}^sa_{ij}^t$, where $a_{ij}^s$ and $a_{ij}^t$ are measures of the spatial and temporal adjacency of the events $i$ and $j$. Further reading on space-time interaction and related problems from other fields can be found in many articles in the literature; \citet{kulldorff99}, \citet{legendre10}, \citet{vanem14} and the references therein being a few examples.

We conclude the section by noting that most of the air pollution studies mentioned earlier in this section concentrated on the pollution issues in different parts of Canada and the USA, while the problem is certainly not limited to this continent only. In fact, there have been only a few good studies that analyze air pollution data from countries in Asia or Africa. A noteworthy paper in this regard is \citet{al06}, who took hierarchical Bayesian approaches to develop a dynamic linear models for air pollutants in Kuwait. They dealt with the temporal and spatial effects independently, defining separate structures for the two processes. 
In general, there is a serious dearth of related papers that focus on the pollution situations in Asia and Africa. And that is another key contribution of this work, as we develop a model and analyze the data for Taiwan. 

The rest of the paper is organized as follows. \Cref{sec:preliminary} 
provides an exploratory analysis of the data under study. The proposed model, 
its properties, and related results are described in \Cref{sec:methods}. 
\Cref{sec:analysis} shows the results of detailed data analysis while \Cref{sec:conclusion} provides some concluding remarks and the scopes of future work.

\setcounter{equation}{0} 

\section{Preliminary Analysis}
\label{sec:preliminary}

\subsection{Data}
\label{subsec:data}

\begin{figure}[!hbt]
\begin{center}
\makeatletter
\def\fps@figure{hbtp}
\makeatother
\includegraphics[width=\textwidth,keepaspectratio]{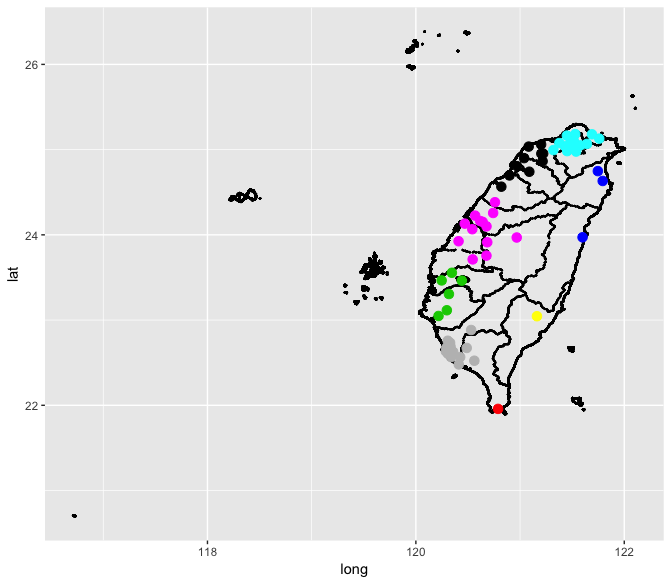}
\caption{Actual map of the locations and clustering for the 66 monitoring stations on Taiwan.}
\label{fig:taiwan-map}
\end{center}
\end{figure}

The $\ppm$ data we analyze are collected from 71 official monitoring stations across Taiwan. However, for five of those stations, we do not have the necessary information on the covariates considered and so, we decide to exclude them from this study. The remaining 66 monitoring stations are irregularly located in space, spread over the whole Island with 
some concentrations around big cities or industrial areas. 
The minimum and maximum of the pairwise distances of these stations are 0.58 kilometers and 366.7 kilometers respectively while the arithmatic mean of the same is 140.7 kilometers. One major talking point of our study is that we consider the issue of space-time interaction as a property of a region, rather than that of every individual station. For that, we divide the data into several clusters, based on the latitude and longitude of the stations. More on this is discussed in the following sections. 
\Cref{fig:taiwan-map} shows the exact locations of the 66 
stations. The concentration of the stations on the west coast is 
understandable because it is the most populated area of the Island. 
Different colors of the map indicate different regions (clusters) 
in our study. 

Turn to the temporal scale of our study. 
The data are obtained on an hourly basis for ten years, starting from January 1, 2006 to 
December 31, 2015. However, the sampling frequencies vary from station to station and for some sites, there are missing values. 
Moreover, as we are mainly interested in identifying space-time interaction for the air pollution data, lower temporal resolution is desired and simpler. Thus, in this analysis, we decide to aggregate the hourly data into weekly averages based on all available measurements within a week. It is worth mentioning that this is a common practice while dealing with monitoring data, as discussed in \citet{smith03}. In order to maintain continuity in the time scale, we consider the whole set of 3652 days (from 2006 to 2015) together and divide it into 522 weeks. Hence, the total number of data points considered in this study is $66\times 522=34452$. Throughout the study, whenever needed, we would use one year's (2015) data, i.e. $52\times 66=3432$ points (approximately the last $10\%$ for each site), for validation purposes. A detailed discussion on this is presented in the following sections.

\begin{figure}[!hbt]
\begin{center}
\makeatletter
\def\fps@figure{hbtp}
\makeatother
\includegraphics[width=\textwidth,keepaspectratio]{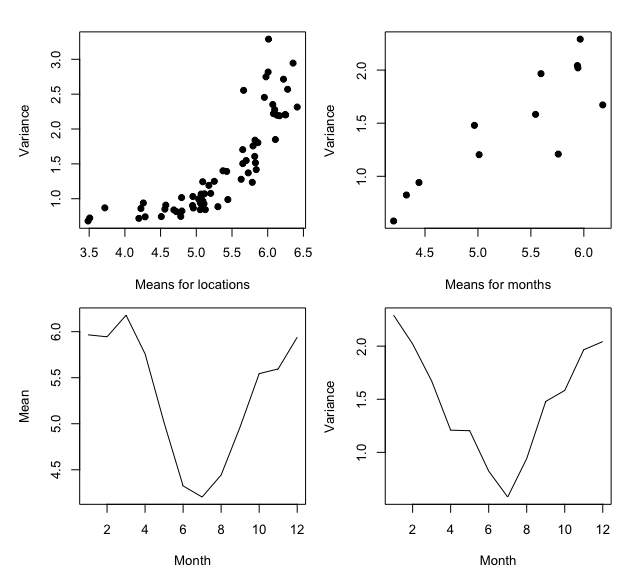}
\caption{(Top left) Sample variances versus means of the weekly $\sqrt{\ppm}$ observations for 66 stations, (Top right) Means versus variances of the $\sqrt{\ppm}$ observations for 12 months, (Bottom left) Means of the $\sqrt{\ppm}$ observations corresponding to different months, (Bottom right) Variances of the $\sqrt{\ppm}$ observations corresponding to different months.}
\label{fig:taiwan-mean-var}
\end{center}
\end{figure}

To begin, we present some exploratory analysis on the data. As has been done in several related studies, we convert the $\ppm$ values to the square root scale and the rest of the study is performed with the transformed data. This type of transformation is a common practice for air pollution data, as discussed in \citet{smith03}. The graph in \Cref{fig:taiwan-mean-var} shows the overall means and the variances of the transformed $\ppm$ observations for different stations and different months. The top left panel describes the variances against the means of the stations. The top right panel shows the same, but for different months in the study. The bottom two plots show the behavior of the means and variances corresponding to different months.

From the top two plots, it is clear that the variance increases 
in a nonlinear manner as the mean increases. On the other hand, the bottom left plot establishes that there is seasonality in the data, which is expected for most related time series problems. Moreover, interestingly, the bottom right plot of the variance corresponding to the months suggests that the variance cannot be assumed to be homoskedastic. In fact, it varies seasonally, and that motivates us to consider a heteroskedastic nature for the error variance in the Gaussian process of the 
proposed model. 

\begin{figure}[!hbt]
\begin{center}
\makeatletter
\def\fps@figure{hbtp}
\makeatother
\includegraphics[width=\textwidth,keepaspectratio]{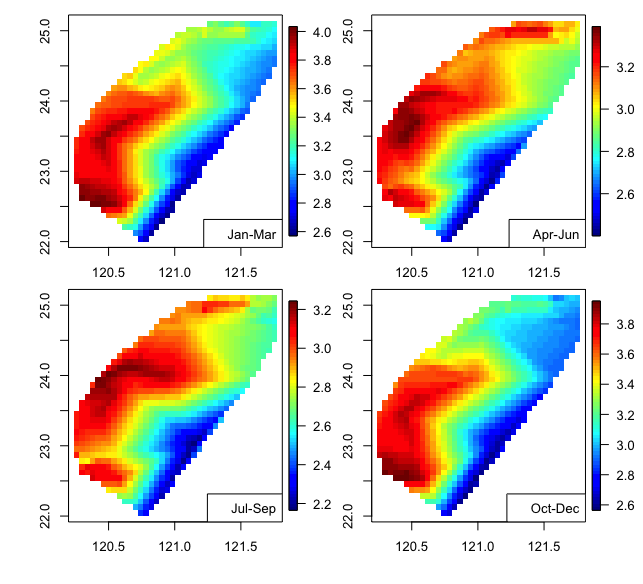}
\caption{Seasonal heat map of the $\ppm$ levels on Taiwan.}
\label{fig:heatmap}
\end{center}
\end{figure}

On the other hand, we also show a heat map (\Cref{fig:heatmap}) of the weekly average of $\ppm$ observations for all the locations. In the map, we show the averages for different seasons. It is evident that the spatial pattern of the weekly averages changes according to seasons, and that motivates us to consider the space-time interaction coefficient in the model, as described in \Cref{sec:methods}. 

\subsection{Covariates used in the study}
\label{subsec:covariates}

For each station, along with the observations of $\ppm$, data were collected on temperature, relative humidity, and wind speed and 
direction. Similar to the air pollution observations, these data were also collected on an hourly basis from January 1, 2006 to December 31, 2015. But, we aggregate them per week and use the representative values as covariates in our study.

\begin{figure}[!hbt]
\begin{center}
\makeatletter
\def\fps@figure{hbtp}
\makeatother
\includegraphics[width=\textwidth,keepaspectratio]{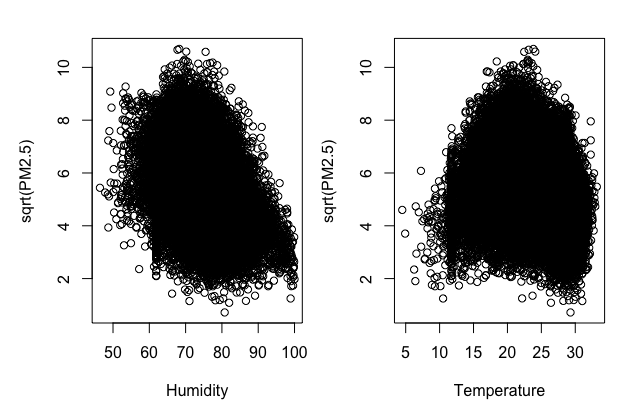}
\caption{Scatter plots of the square roots of the $\ppm$ observations, with respect to relative humidity (left) and temperature (right)}
\label{fig:taiwan-covariates}
\end{center}
\end{figure}

At first, we examine the relationship of the pollution data with the relative humidity and the temperature. The scatter plots of the square root transformations of $\ppm$ against these covariates are shown in \Cref{fig:taiwan-covariates}. From a quick glance, it seems that the air pollution is not so significantly affected by the temperature, but is dependent on the humidity. It shows a slight decrease in pollution with the increase in the relative humidity. The Spearman correlation coefficient between transformed $\ppm$ and humidity was found out to be $-0.326$, while the same for the temperature was $-0.254$. 

\begin{figure}[!hbt]
\begin{center}
\makeatletter
\def\fps@figure{hbtp}
\makeatother
\includegraphics[width=\textwidth,keepaspectratio]{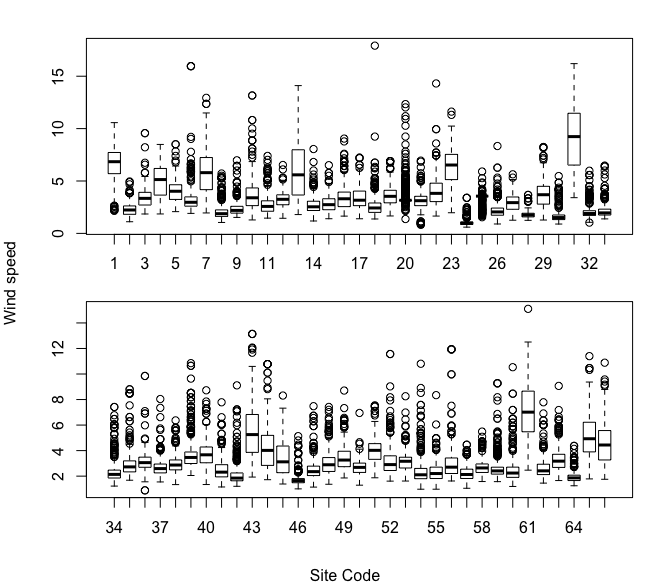}
\caption{Boxplots of the wind speed for 66 stations}
\label{fig:boxplot-wind}
\end{center}
\end{figure}

The wind speed is another variable that may have some effects on the air pollution. As before, the data on wind speed and wind direction were available on an hourly basis. It was noted that the behavior of the wind speed and direction varied widely for different locations. To prove this point, boxplots of the daily wind speed, corresponding to all the locations, are displayed in \Cref{fig:boxplot-wind}. 

The boxplots shows clearly that the ranges of the wind speed vary markedly for different locations, while the means are not so much. Naturally, in order to account for the effect of the wind, 
unlike the previous covariates, 
it will be wise not to take the weekly average. Instead, we decided to use the maximum daily wind speed for every week. It is also worth mentioning that for most weeks, the maximum wind speed occurred on same day in all the locations, which again justifies why the maximum in lieu of the average is a good measure in this regard. The sample 
correlation coefficient between transformed $\ppm$ and the maximum wind speeds is $-0.146$.


\setcounter{equation}{0} 

\section{Methods}
\label{sec:methods}

\subsection{The proposed model}
\label{subsec:proposed-model}

The proposed model is in some sense inspired by the one studied by \citet{sahu06}. The key feature of our model is that it accounts for possible space-time interactions in the air-pollution data. We describe the model in a general setup, and discuss the particular case of Taiwan data in \Cref{subsec:ourmodel}.

Suppose the data are collected for $n$ locations and over $T$ consecutive time points. Let us denote the square root of the $\ppm$ at time $t_j$ and location $s_i$ by $Z(s_i,t_j)$, for $i=1, \hdots, n$ and $j=1, \hdots, T$. Then, we assume that the overall mean values for different locations are different, and we subtract the location-wise overall means from the actual values of the transformed $\ppm$ values to obtain the mean-adjusted numbers. These mean-adjusted values are denoted by $Y(s_i,t_j)$. For convenience, we drop the subscripts whenever not needed.  In the proposed model, we consider the following hierarchical structure:
\begin{equation}
\label{eqn:hierarchical_model1}
Y(s,t) := U(s,t) + \epsilon(s,t),
\end{equation}
where $U(s,t)$ describes a spatio-temporal process and $\epsilon(s,t)$ denotes a white noise process, to account for the measurement errors. We assume the white noise process to follow heteroskedastic $N(0,\sigma_i^2)$ distributions independently, where $\sigma_i^2$ is chosen according to different seasons. $U(s,t)$, on the other hand, assumes the following structure:
\begin{equation}
\label{eqn:hierarchical_model2}
U(s,t) := \mu(s,t) + v(s,t), 
\end{equation}
where $\mu(s,t)$ stands for the mean of the $U(s,t)$ process and $v(s,t)$ denotes a zero-mean spatio-temporal process.

The aforementioned mean function is considered to be an additive combination of the effects of the covariates, the seasonal effects, and a spatio-temporal interaction effect. In order to capture the effects of available covariate information, we  consider a term of the form $B\alpha$ where $B$ is the design matrix for $p$ covariates. The population densities, temperature, humidity, number of factories, number of cars, etc can be taken as the covariates for the analysis. The seasonal variation is captured by introducing indicators for different seasons. Throughout this study, we have considered monthly indicators, but the method can be used for other cases too. In general, let us use $J$ to denote the number of seasons in a year.

The spatio-temporal effect we use in the mean structure allows us to test for space-time interaction, if any. In principle, the effect is described as a linear function of the form $\gamma_0+\gamma_st$, thereby allowing us to test if the $\gamma_s$ values are same across different locations. However, this would increase the complexity and computational burden dramatically, if we have a lot of sites. To address this issue, we make a logical assumption that the coefficients $\gamma_s$ for locations which are close to each other are the same, and that it should be a characteristic of a region rather a station. 
So, based on the coordinates of the stations, we divide 
the stations into clusters first and employ $\gamma_k$ for all stations in the $k$th cluster. The number of clusters we choose is based on the number of locations we have. If $n$, the number of locations, is 5 or less, we do not use clustering. But for $n>5$, we will use $[\sqrt n]$ clusters ($[\cdot]$ denotes the greatest integer function) to divide the locations into different regions. In practice, 
we use the $k$-means clustering method, based on the latitude and longitude, 
for this purpose. 

Further, we propose to use an intercept-free model, and so, along with the response variable, we also subtract the overall means of the covariates from the respective observations. Thus, the mean structure of the process can be described as:
\begin{equation}
\label{eqn:model_mean}
\mu(s,t) = c(s,t)'\alpha + \sum_{j=2}^{J} \beta_j \ m(t,j) + \sum_{k=1}^K \gamma_k \ r(s,k) t.
\end{equation}
In the above, $c(s,t)$ is a column vector with the values of the covariates and $m(t,j)$ and $r(s,k)$ are the indicators for the season and location region, respectively, where $J$ and $K$ are the seasonality and the number of regions. 
That is, $m(t,j)=1$ if time $t$ is in the $j$th season and 0 otherwise. Similarly, $r(s,k)=1$ if location $s$ is in $k$th region and 0 otherwise. For identifiability purposes, we would take $\beta_1=0$. Note that because of using mean-adjusted values for the response and the covariates, we do not have any intercept term in the model. On the other hand, we would scale down the time points ($t$) to equi-spaced points in the interval $[0,1]$.

The term $v(s,t)$ in \Cref{eqn:hierarchical_model2} can be treated as a spatially varying temporal trend. Averaging over different sites in a region, we can get the adjustment to the regional trends and averaging over time can help us obtain the adjustment at the temporal scale. For convenience, we consider a separable structure for the covariance of this process. Moreover, we assume that the locations in different cluster are independent of each other. In particular, the covariance between $v(s_1,t_1)$ and $v(s_2,t_2)$, when $s_1,s_2$ are in the same cluster, is taken as a product of the spatial dependence and temporal dependence and we write it as 
\begin{equation}
\label{eqn':spatio-temporal-covariance}
\cov(v(s_1,t_1),v(s_2,t_2)) = \sigma_v^2 \ \rho(\norm{s_1-s_2},\phi_s)\cdot\rho(\norm{t_1-t_2},\phi_t) \cdot\I\{s_1,s_2 \in \mathcal{C}_k\}, 
\end{equation}
where $\rho(x,d)$ denotes the exponential covariance function $e^{-dx}$, and $\mathcal{C}_k$ denotes a particular cluster. The distance functions $\norm{s_1-s_2}$ or $\norm{t_1-t_2}$ are taken as the Euclidean distance of the two points. 

Throughout this paper, $Y$ would denote the vector of $N=nT$ data points, arranged according to clusters at first, time points next, and sites then. Thus, if $\{s_1,s_2\}$ form a cluster, then the first few observations will be $Y(s_1,t_1), Y(s_2,t_1), Y(s_1,t_2), Y(s_2,t_2)$, etc. 
The vector $v=(v(s_i,t_j))$ is formed in a similar way. We denote the full covariance matrix of $v$ by $\Sigma_v$. Note that $\Sigma_v$ is a block diagonal matrix, where each block corresponds to the covariance matrix of a cluster of locations, and is of the form $\sigma_v^2(\Sigma_t \otimes \Sigma_s)$ such that $\Sigma_t(i,j)=\rho(\norm{t_i-t_j},\phi_t)$ and $\Sigma_s(i,j)=\rho(\norm{s_i-s_j},\phi_s)$. Further, note that when the sampling design considers equally-spaced time points, the common spacing being $d_t$, one can write $\Sigma_t(i,j)=\psi^{\abs{i-j}}$, where $\psi=e^{-\phi_td_t}$.

On the other hand, as mentioned before, we entertain a heteroskedastic error function for $\epsilon(s,t)$. The exploratory analysis suggests that the variances are different for different seasons, and so, we assume $\epsilon(s,t) \sim N(0,\sigma^2_{m(t)})$, where $m(t)$ denotes the season the time $t$ is in. If we use $\epsilon$ to denote the vector of $\epsilon(s,t)$, arranged similarly as $Y$ and $w$, then the above discussion implies that $\epsilon\sim N(0,\sigma^2 D)$, where $D$ is a diagonal matrix such that the diagonal element corresponding to $\epsilon(s,t)$ is $\sigma^2_{m(t)}/\sigma^2$. 
We denote these parameters by $\tau_1^2, \hdots, \tau_J^2$, where $\tau_i^2$ is the variance parameters associated with the $i$th season.  Also, we set $\tau_1^2=1$ to avoid any potential identifiability problem.

Now, to write the full model in a vector-matrix notation, recall that $Y$ is the observed 
vector  of dependent varaible, and $v$ and $\epsilon$ denote the corresponding vectors of the zero mean spatio-temporal process and the Gaussian error process, respectively. We can write the mean function as the sum of $c(s,t)'\alpha$, $m_t'(\beta_j)$ and $r_{st}'(\gamma_k)$, where $m_t$ and $r_{st}$ are the column vectors corresponding to the parameter vectors $(\beta_j)_{2\le j\le J}$ and $(\gamma_k)_{1\le k\le K}$, respectively. Then, denoting the vector of all the parameters by $\theta$ and letting $X$ be a design matrix such that each row is of the form $X(s,t)'= (c(s,t)', m_t', r_{st}')$, the model can be written as:
\begin{align}
\label{eqn:full-model}
Y &= X \theta + v+ \epsilon, \\
\text{where } \ v &\sim N(0,\sigma_v^2\Sigma_v), \nonumber \\
\text{and } \ \epsilon &\sim N(0,\sigma^2D). \nonumber
\end{align}

It is evident that there are $(p+J+K)$ components in the parameter vector $\theta$ if we consider $p+1$ covariates and $K$ regions for the locations. On the other hand, we write the two variance components $\sigma^2, \sigma_v^2$ to be equal. Here, the estimate of $\sigma^2$ would give us an idea about the variance explained by the spatio-temporal process while the estimates of the diagonal elements of $D$ tell us how much is explained by the pure error process.

Finally, the best estimates for $\phi_s$ and $\phi_t$  are obtained using a cross validation scheme. The validation scheme considers prediction for all the sites at some time points and obtains the mean squared error for those predictions. In this study, for every site, we take the first 80\% of the time points under consideration (e.g. in case of weekly data for 10 years, we would consider the first 8 years or 418 weeks) for model fitting and then make predictions for the last 20\% (104 weeks for the aforementioned data) to see which combination of $(\phi_s,\phi_t)$ would work the best. The possible choices for $\phi_s$ used in the study were (0.001, 0.005, 0.01, 0.05, 0.1, 0.3, 0.5) while the choices for $\phi_t$ were (0.5, 0.75, 1, 1.5). To find out the optimal choice, we find the combination with the smallest mean squared error of the predictions, which is calculated as follows. For each site $s_i$, let us denote the validation time points by $t_1, \hdots, t_b$ and the predictions by $\hat Y(s_i,t_j)$ for $i=1, \hdots, n$; $j=1, \hdots, b$. Then, for each of the $5\times 4=20$ combinations, the prediction mean squared error is computed as:
\begin{equation}
\label{eqn:validation-mse}
\text{MSE} = \frac{1}{nb}\sum_{i=1}^n\sum_{j=1}^{b} \{ Y(s_i,t_j) - \hat Y(s_i,t_j) \}^2.
\end{equation}
The prediction procedure is discussed in \Cref{subsec:prediction}.

\subsection{Testing for interaction and parameter estimation}
\label{subsec:testing}

The model described in \Cref{eqn:full-model} allows us to test for space-time interaction. The model includes no interaction term if $\gamma_1=\hdots=\gamma_K=0$. We employ the 
Lagrange multiplier (LM) test for this purpose. Recall that the main advantage of the LM test or the score test is that it, unlike Wald test or likelihood ratio test, does not require an estimate of the information under the alternative hypothesis or unconstrained maximum likelihood. The LM test uses only the assumptions in the null hypothesis to get the maximum likelihood estimates and then calculates the value of the test statistic (which follows a chi-squared distribution with appropriate degrees of freedom) to make a decision.

Once we perform the LM test, we can make a decision about the model to use. We use the 
full model (\ref{eqn:full-model}) if the decision is to reject the null hypothesis that there is no space-time interaction. 

To estimate the parameters, we employ the generalized least squares techniques, with little modifications. Observe that the proposed model can be thought of in the form $Y=X\theta+\varepsilon$, where $\varepsilon \sim N(0,\sigma^2\Omega)$ such that $\Omega=\Sigma_v+D$. This is in the setup of a generalized least squares problem. Further, note that the number of unknown parameters in the error covariance matrix $\Omega$ is $J+2$, namely $\sigma^2, \tau_2^2, \hdots, \tau_J^2; \phi_s, \phi_t$. As we decided to get the optimal choices for the last two parameters by a cross-validation procedure, that leaves us the task of estimating $J$ additional variance parameters from the model.

From a practical point of view, it is more important to identify the cases where the air pollution is hazardous for health and environment, rather than the ones when the situation is less harmful. According to the standards set by the EPA, the average for the fine particulate matters should not exceed $35\ \unit$. So, while minimizing the squared sum of residuals, 
we put more weight on the cases where the actual $\ppm$ values are more than $35$. Thus, 
the loss function that we want to minimize is of the form
\begin{equation}
\label{eqn:loss}
L = \sum_{i=1}^n\sum_{j=1}^T w(s_i,t_j)  \hat e(s_i,t_j)^2.
\end{equation}
Here, $w(s_i,t_j)$ should be taken higher for $Z(s_i,t_j)\ge \sqrt{35}$. We take it to be $(1+2/\log N)$ for $Z(s_i,t_j)\ge \sqrt{35}$ and $(1-2/\log N)$ otherwise, where $N=nT$. An attractive feature of this is that the weights will approach 1 as $N\to\infty$, thereby establishing that the importance will be approximately equal on all observations when the sample size is huge. $\hat e(s_i,t_j)$, on the other hand, is the standardized residual for location $s_i$ and time $t_j$. Note that it can be written as $\hat\Omega^{-1/2}\hat\varepsilon/\hat\sigma = \hat\Omega^{-1/2}(Y-X\hat\theta)/\hat\sigma$. Thus, assuming that $\hat\Omega$ and $\hat\sigma$ are known to us, we can say that minimizing $L$ is equivalent to minimizing $(Y-X \theta)'\hat\Omega^{-1/2}W\hat\Omega^{-1/2}(Y-X\theta)$ with respect to $\theta$. Here $W$ is a diagonal matrix with the diagonal elements being the same as $w(s_i,t_j)$. 
It is evident that the minimizer is $\hat\theta = (X^TVX)^{-1}(X^TVY)$, where $V=\hat\Omega^{-1/2}W\hat\Omega^{-1/2}$. In light of the above discussion, we propose to use the following procedure to estimate the parameters in the model.

{\bf Parameter estimation procedure:}
\begin{enumerate}
\item Set $\hat\tau_j^2=1$ for $j=1, \hdots, J$.
\item Evaluate $\hat\Omega$ and $V=\hat\Omega^{-1/2}W\hat\Omega^{-1/2}$.
\item Compute $\hat\theta = (X^TVX)^{-1}(X^TVY)$ and set $\hat\varepsilon = Y-X\hat\theta$.
\item Compute $\hat\sigma^2=\hat\varepsilon^T\hat\Omega^{-1}\hat\varepsilon/N$, where $N=nT$ is the total number of observations.
\item Let $\hat\varepsilon_j$ be the error corresponding to the $j$th season, for $j=1, \hdots, J$.
\item For $j=2, \hdots, J$, note that $\varepsilon_j\sim N(0,\sigma^2(\Sigma_v^{(j)}+\tau_j^2I))$, where $\Sigma_v^{(j)}$ is the submatrix of $\Sigma_v$ corresponding to the $j$th month. 
\item Use optimization methods to compute the MLE $\hat\tau_j^2$ using the above.
\item Repeat steps 2 to 7 until convergence.
\end{enumerate}

In the above procedure, in order to reduce the computational burden, we exploit the block-diagonal structure of the matrix $\Omega$. Note that, if each block of $\Omega$ is denoted by $B_i$, then $\Omega^{-1/2}$ can be written as a block diagonal matrix, where the blocks are $B_i^{-1/2}$. Using this and writing $W$ and $X$ accordingly in the above steps, we substantially reduce the computational burden in the estimation.

The following theorem describes the asymptotic properties of the above estimators. Proof of the theorem is provided in \Cref{sec:appendix}.

\begin{theorem}
\label{thm:asymptotics}
The estimate $\hat\theta$ obtained from the above procedure is consistent, in the sense that as the number of locations ($n$) and the number of time points ($T$) approach infinity, $\hat\theta \to \theta$ in probability. 
Furthermore, $\sqrt{nT}(\hat\theta-\theta) \to N(0,\sigma^2Q^{-1})$, where $Q$ is the limit of $(X\dash\Omega^{-1}X)/nT$ as $n,T\to\infty$.
\end{theorem}

\subsection{Future prediction}
\label{subsec:prediction}

To make a new prediction for site $s\dash$ at time $t\dash$, we use the parameter estimates  obtained from the method mentioned in the previous section. Let us use $X(s\dash, t\dash)$ to denote the new set of covariate vectors, similar to what we did in \Cref{subsec:proposed-model}. Then, it can be said that $\hat Y(s\dash,t\dash) = X(s\dash, t\dash)'\theta+\varepsilon(s\dash,t\dash)$. However, in view of the fact that $\Omega$ is not a constant multiple of the identity matrix, we cannot simply assume that the prediction error is going to be independent of the sample disturbances. The prediction procedure has to take the dependence into account.

If $\varepsilon$ is the error vector corresponding to the original data, let us denote $\cov[\varepsilon(s\dash,t\dash),\varepsilon]$ by $w$, which is going to be a $nT$-dimensional column vector. It is evident that $w$ and $\Omega$ depend on the values of $\phi_s,\phi_t,\sigma^2$ and $\tau_j^2$ for $j=1, \hdots, J$. We use the estimates of these parameters to get $\hat w$ and $\hat\Omega$. Then, following \citet{goldberger62}, we can say that the best linear unbiased predictor is 
\begin{equation}
\label{eqn:prediction}
\hat Y(s\dash,t\dash) = X(s\dash,t\dash)'\hat\theta + \frac{1}{\hat\sigma^2}\cdot\hat w\dash\hat\Omega^{-1}(Y-X\hat\theta).
\end{equation}
Finally, we shift the location and transform the predictions to get the estimate of the actual air pollution measurement.

\section{Detailed Analysis}
\label{sec:analysis}

\subsection{Simulation studies}

To begin, we present some simulation studies to show that our methods are capable of capturing the space-time interaction that might be present in the empirical application we are 
interested in. 

We perform the simulation experiment in two stages - first with a linear space-time interaction term and then with a non-linear one. Throughout this study, we consider weekly average of the air pollution data. And then, for different time intervals, we evaluated the type-I error and the power for different values of $n$. In the simulation study, we considered four different values of $T$: 52 (1 year), 104 (2 years), and 261 (5 years). We worked with different numbers of locations ($n= 10, 20, 50$) to understand how the proposed method performs as we have more data. For $n$ locations, the $xy$-coordinates were generated randomly from a $(0,n^2)\times (0,n^2)$ grid. The values of the variance parameters $\sigma^2,\tau_j^2$ (see \Cref{subsec:proposed-model}) were generated from an inverse-gamma distribution with parameters $(4,3)$. All the coefficients in the model were simulated from independent normal distributions with mean 0 and standard deviation 1. Finally, we used $\phi_s=0.1,\phi_t=0.75$ for the covariance matrix of the spatio-temporal process $w(s,t)$.


For the case of linear space-time interaction, the observations (square root of the $\ppm$ data) were obtained from the model (\ref{eqn:full-model}). However, for the nonlinear case, we replaced the linear term with the quadratic term $(t/T)^2$, and assumed that the space-time interaction coefficients are negative. This type of interaction is common, and usually the coefficients are not big in magnitude. We wanted to observe whether a linear approximation works well to capture this type of interaction as well. The results for different cases under the linear time dependence are displayed in \Cref{tab:simulation-large}, while those for the nonlinear time dependence are shown in \Cref{tab:simulation-large-NL}. The results reported are based on 500 iterations of simulation.

\begin{table}[!hbt]
\caption{Results for Different Cases, When the Dependence on Time is Linear. 5\% critical 
values and 500 iterations are used.}
\centering
\label{tab:simulation-large}
\vspace{0.1in}
\begin{tabular}{|c|ccc|ccc|}
\hline
 & \multicolumn{3}{|c|}{Type-I error} & \multicolumn{3}{|c|}{Power}\\
\hline
Number of locations & 10 & 20 & 50 & 10 & 20 & 50 \\
\hline\hline
Weekly (1 year)     & 0.052  & 0.050  & 0.048  & 0.224  & 0.512 & 0.786 \\
Weekly (2 years)    & 0.056  & 0.058  & 0.042  & 0.312  & 0.578 & 0.820 \\
Weekly (5 years)    & 0.054  & 0.044  & 0.046  & 0.428  & 0.702 & 0.868 \\
\hline
\end{tabular}
\end{table}

While the type I error remains under control across all scenarios, the power improves significantly for 20 or more locations and this is true both for the linear dependence and for the nonlinear dependence. The power also increases with sample size for all cases. The results confirm that the proposed testing procedure does not have large size distortions and, as expected, fares well for larger sample sizes.

\begin{table}[!hbt]
\caption{Results for Different Cases, When the Time-Dependence is Nonlinear. 5\% critical 
values and 500 iterations are used}
\centering
\label{tab:simulation-large-NL}
\vspace{0.1in}
\begin{tabular}{|c|ccc|ccc|}
\hline
 & \multicolumn{3}{|c|}{Type I error} & \multicolumn{3}{|c|}{Power}\\
\hline
Number of locations & 10 & 20 & 50 & 10 & 20 & 50 \\
\hline\hline
Weekly (1 year)     & 0.058  & 0.046  & 0.046  & 0.232  & 0.394 & 0.678 \\
Weekly (2 years)    & 0.048  & 0.042  & 0.062  & 0.268  & 0.478 & 0.714 \\
Weekly (5 years)    & 0.058  & 0.050  & 0.066  & 0.358  & 0.500 & 0.794 \\
\hline
\end{tabular}
\end{table}

\begin{figure}[!hbt]
\begin{center}
\makeatletter
\def\fps@figure{hbtp}
\makeatother
\includegraphics[width=\textwidth,keepaspectratio]{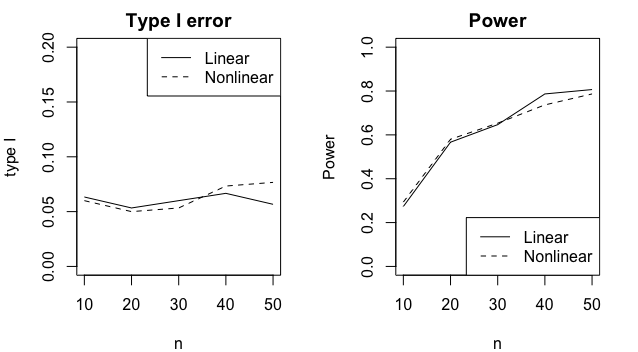}
\caption{Type-I error (on left) and power (on right) for linear (solid) and nonlinear (dotted) cases, corresponding to different number of locations. $T=157$ (3 years) and 500 iterations are used for all cases.}
\label{fig:simulation-size-power}
\end{center}
\end{figure}

We further extended our simulation studies to see how the proposed 
method behaves for larger data. To understand this, we concentrated on weekly data from 3 years (157 observations for each site) and took different values for $n$, ranging from 10 to 50. The results of 
500 iterations for different cases are shown in \Cref{fig:simulation-size-power}. It is evident that, when the dependence on time is linear, the proposed testing procedure fares nicely for large number of sites. The testing procedure also works well for the nonlinear case as its power also increases,  reaching about 0.8 when the number of locations is 50.

\begin{figure}[!hbt]
\begin{center}
\makeatletter
\def\fps@figure{hbtp}
\makeatother
\includegraphics[width=\textwidth,keepaspectratio]{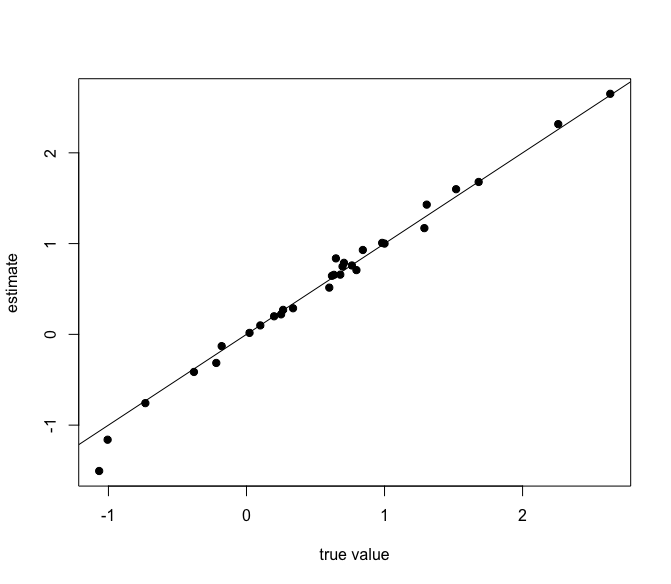}
\caption{True value and estimated value for the parameters in the model, where data is generated for 20 locations and 5 years}
\label{fig:simulation-estimates}
\end{center}
\end{figure}

Finally, we present a particular case to show that the parameter estimation process indeed works well to identify the effect of the factors. In this example, we used 20 different locations, two covariates (humidity and temperature) and simulated data from our model to get weekly averages for 5 years (divided in 12 seasons). Thus, the number of observations in the data were $20\times 261=5220$. We then estimated the parameters using the procedure described in \Cref{subsec:testing}. In \Cref{fig:simulation-estimates}, the true values and the estimates of all the parameters in the model are plotted. We can see that most points lie along the line $y=x$ (displayed in the figure), thereby showing that the estimates are not too different from the true values. This study confirms that the proposed iterative estimation procedure is reliable.

\subsection{Model selection}
\label{subsec:ourmodel}

The first task to implement the proposed method is to choose proper decay parameters and to identify if there exists any space-time interaction. For the decay parameters, we searched in a two-dimensional array to find out which combination gives the least mean-squared error (refer to \Cref{eqn:validation-mse}). The choices for $\phi_s$ were $(0.001, 0.005, 0.01, 0.05, 0.1, 0.3, 0.5)$ while the same for $\phi_t$ were $(0.5, 0.75, 1, 1.5)$. We used 80\% of the data as our training data and the rest 20\% were used for validation purposes in this regard.

We found that the combination of $\phi_s=0.01$ and $\phi_t=1$ was the best one for the data at hand. To put it into the perspective of the actual spatial and temporal scale, we can say that these choices correspond to a significant correlation in an approximate range of 300 kilometers and in a  time span of 3 weeks, respectively. On the other hand, the LM test returned a $p$-value less than the level of significance (0.05), establishing that a space-time interaction effect is indeed present in the data. Thus, the model we decide to fit for our empirical analysis is the same as \Cref{eqn:full-model}, which is shown below for ease in reference.
\begin{equation}
Y(s,t) =  c(s,t)^T\alpha + \sum_{j=2}^{12} m(t,j)\beta_j + \sum_{k=1}^K r(s,k)\ \gamma_kt + v(s,t) +  \epsilon(s,t), \nonumber
\end{equation}
for $s=1, \hdots, n;\ t=1, \hdots, T.$ Recall that $c(s,t)$ describes the covariates in the study, and we use relative humidity, temperature, and wind speed in the analysis. Furthermore, $T=522$, $n=66$ and thus, the number of clusters ($K$) is taken to be 8. 

\subsection{Parameter estimates}

Next, we fitted the above model to Taiwan data to obtain parameter estimates. In this 
particular application, we have 22 parameters in the mean structure and 12 more in the variance structure.  In what follows, we discuss these estimates step-by-step. \Cref{tab:estimates-covariates} shows the coefficient estimates corresponding to the covariates and the estimates of the seasonal effects. 

\begin{table}[!hbt]
\caption{Estimates of the Overall Location Mean Parameters}
\centering
\label{tab:estimates-covariates}
\vspace{0.1in}
\begin{tabular}{lccc}
\hline
Parameter & Estimate & Standard error & Confidence interval \\
\hline
$\alpha_1$ (Humidity) 	& $-0.0318$ & 0.0008 & $(-0.0334,-0.0302)$ \\
$\alpha_2$ (Temperature) & $0.0203$ & 0.0024 & $(0.0156,0.0250)$ \\
$\alpha_3$ (Wind speed) 	& $-0.0779$ & 0.0038 & $(-0.0853,-0.0705)$ \\
\hline
$\beta_2$ (February) 	& $0.0100$ & 0.0257 & $(-0.0403,0.0603)$ \\
$\beta_3$ (March) 		& $0.0682$ & 0.0267 & $(0.0159,0.1206)$ \\
$\beta_4$ (April) 		& $-0.3526$ & 0.0296 & $(-0.4105,-0.2946)$ \\
$\beta_5$ (May) 		& $-1.1230$ & 0.0342 & $(-1.1900,-1.0561)$ \\
$\beta_6$ (June) 		& $-1.7756$ & 0.0378 & $(-1.8497,-1.7016)$ \\
$\beta_7$ (July) 		& $-1.9105$ & 0.0395 & $(-1.9879,-1.8331)$ \\
$\beta_8$ (August) 		& $-1.5696$ & 0.0390 & $(-1.6460,-1.4932)$ \\
$\beta_9$ (September) 	& $-0.9832$ & 0.0375 & $(-1.0568,-0.9096)$ \\
$\beta_{10}$ (October) 	& $-0.4402$ & 0.0330 & $(-0.5049,-0.3756)$ \\
$\beta_{11}$ (November) 	& $-0.3213$ & 0.0294 & $(-0.3790,-0.2636)$ \\
$\beta_{12}$ (December) 	& $0.0494$ & 0.0257 & $(-0.0010,0.0998)$ \\
\hline
\end{tabular}
\end{table}
From the table, it is seen that almost all of the estimates show a significant effect.  The $\ppm$ is higher whenever the humidity is less and the temperature is higher. Also,  as expected, a higher wind-speed reduces the amount of pollution significantly. On the other hand, there exists a strong seasonal pattern. The winter months (from December to February) does not show a significant change in the pollution. March shows a significant increase whereas the $\ppm$ decreases over the summer months,  especially between June and August. This is understandable given the geographical location  and climate pattern of Taiwan. It is more likely to rain over the summer in Taiwan and  the wind tends to be from the south-east and the speed can be high with possible strong  typhoons. On the other hand, winter months tend to be dry with wind from north-west in Taiwan. 

\begin{figure}[!hbt]
\begin{center}
\makeatletter
\def\fps@figure{hbtp}
\makeatother
\includegraphics[width=\textwidth,keepaspectratio]{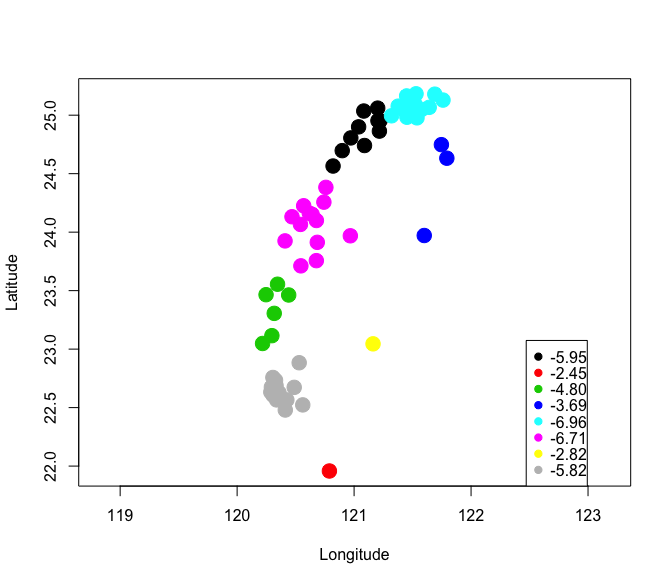}
\caption{Estimates of the space-time interaction terms for different regions}
\label{fig:estimates-spacetime}
\end{center}
\end{figure}

We then plotted the space-time interaction coefficients to see how they spread across the whole Island. Figure \ref{fig:estimates-spacetime} shows these estimates on a spatial scale. It was found that all of the regions showed significantly negative estimates for the space-time interaction coefficients, the extent of interaction being different from region to region. In absolute sense, it is the least for stations Hengchun (denoted by red) and Guanshan (denoted by yellow) while it is the most for stations around Taipei (light blue), Taichung (magenta), 
Taoyung (black), and Kaohsiung (grey) of Taiwan. This is also understandable as Hengchun is at the southern tip of Taiwan and Guanshan is less populated and without any heavy industry. 
On other other hand, the most significant stations are around the four largest cities in Taiwan. 

\begin{figure}[!hbt]
\begin{center}
\makeatletter
\def\fps@figure{hbtp}
\makeatother
\includegraphics[width=\textwidth,keepaspectratio]{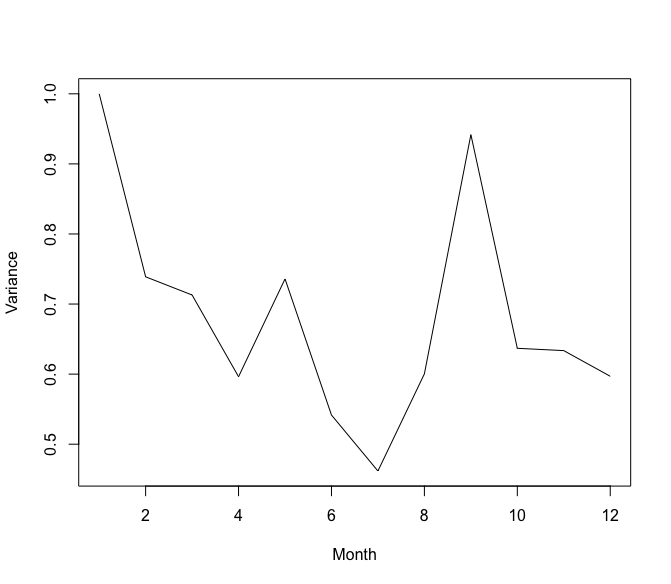}
\caption{Estimates of the variance term for different months}
\label{fig:estimates-variance}
\end{center}
\end{figure}

The estimates of the variance parameters for different months are shown in \Cref{fig:estimates-variance}. We can see that the estimates show an oscillatory behavior across different months, and the magnitudes of the variances range between 0.4 and 1. These estimates can explain the extent of the heteroskedasticity in the white noise process in our model. So far as the spatio-temporal process is concerned, its variance $\sigma^2$ was estimated at 0.8462, which is more than most variance terms of the white noise. This indicates that the white noise process can explain less variability in the data, and that the spatio-temporal process is more significant in this aspect.
 
\subsection{Model diagnostics}

\begin{figure}[!hbt]
\begin{center}
\makeatletter
\def\fps@figure{hbtp}
\makeatother
\includegraphics[width=\textwidth,keepaspectratio]{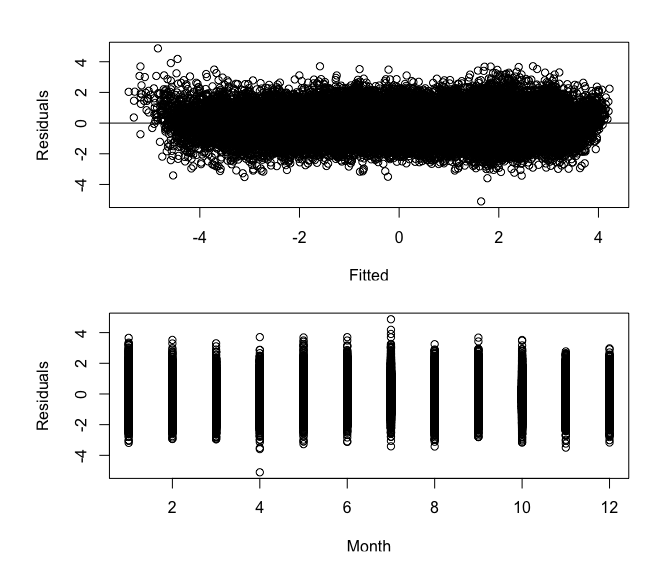}
\caption{(Top) Standardized residuals are plotted against fitted values; (Bottom) Standardized residuals are plotted corresponding to different months}
\label{fig:residuals-fitted}
\end{center}
\end{figure}

In this section, we present some residual plots to show that the  proposed model fares well to capture the effects of Taiwan $\ppm$ data.  We evaluate the residuals after fitting the model and plot them using standard statistical  procedures. First, the top panel in \Cref{fig:residuals-fitted} shows that there is no particular pattern in the residuals, thereby establishing that the residuals can be assumed to be uncorrelated. Next, the plot of the residuals corresponding to different months are presented in the bottom panel of the same figure. It shows that the variances are more or less evenly distributed across the months, thereby showing that the issue of heteroskedasticity has been taken care of. 

\begin{figure}[!hbt]
\begin{center}
\makeatletter
\def\fps@figure{hbtp}
\makeatother
\includegraphics[width=\textwidth,keepaspectratio]{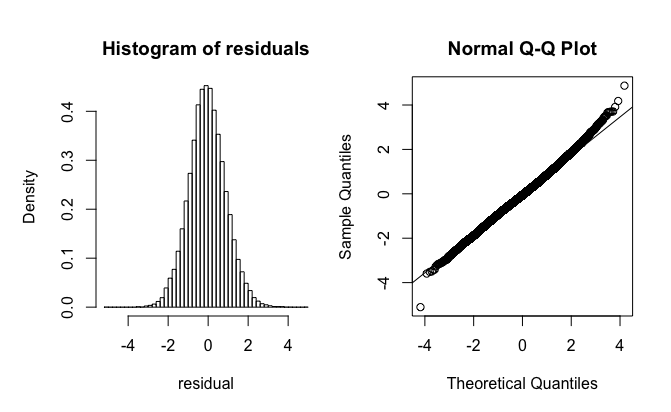}
\caption{(Left) Histogram and (Right) QQ plot of the standardized residuals}
\label{fig:residuals-normality}
\end{center}
\end{figure}

Furthermore, the left panel of \Cref{fig:residuals-normality} shows that the histogram is approximately normally shaped, while the QQ plot presented in the right panel of the same figure corroborates that as well. Consequently, the introduction of heteroskedasticity and the spatio-temporal process works well for the data under study. 

Finally, we check the prediction abilities of the proposed model, using cross validation techniques. For this, we use 90\% of the available data (from 2006 to 2014, for all the stations) and predict the $\ppm$ levels for the year 2015 for all the stations. In order to evaluate how good the predictions are, we have calculated the root mean squared error for the transformed $\ppm$ concentrations, and it is approximately 1.712. In the original scale, the same was found out to be around 13.089 units. In order to understand the effect of the space-time interaction term in our model, we also measured the prediction abilities of the same model, but without the interaction component. In that case, the root mean squared error was found out to be approximately 2.448 in the transformed scale, and approximately 31.341 units in the original scale. To put it into perspective, in the transformed data, this is about 43\% more (about 140\% more in the original scale) than the root mean squared error for the proposed model.

\section{Conclusion}
\label{sec:conclusion}

In this paper, we have developed a new spatio-temporal modeling technique, with an aim to identify the space-time interaction. The simulation studies and the data analysis confirm that the method performs well. In particular, we have showed that the estimates obtained by the proposed method are consistent, and have used standard diagnostic techniques to establish that the model assumptions are reasonable. This modeling technique can successfully detect and estimate the space-time interaction for air pollution data. Further, because of the  weighting scheme we use in the method, it has the potential to predict higher level of pollution with more precision. This is going to be useful 
from a practical point of view.

We finish with some notes on future studies. An important potential future direction of this work is to consider a more generalized framework in the spatio-temporal process. In particular, we consider a separable structure for the spatial and temporal dependence, and that condition can be relaxed to address a more general setup. Moreover, in the aforementioned data analysis example, it was found that maximum wind speed at every location plays an important role in the air pollution, while there is significant space-time interaction effect as well. Combined, they raise an important question - how much effect does the wind flux, calculated from the wind speed, wind direction and the coordinates of different locations, have on the pollution? In order to address that, it will be necessary to know about the physical behavior of the wind, and that process can be incorporated in the model to develop more efficient techniques. 

Another important aspect of this topic is to identify the effect of air pollution on human life. Pollution is responsible for many respiratory and cardiovascular diseases, and it is immensely important to use appropriate modeling techniques for such problems. It has the potential to impact health sciences significantly.

\section{Appendix}
\label{sec:appendix}

\begin{proof}[{\bf Proof of \Cref{thm:asymptotics}}]

Note that the set-up of our problem is similar to a generalized least squares (GLS) problem, where $Y=X\theta+\varepsilon$, such that $\varepsilon\sim N(0,\sigma^2\Omega)$. Following our previous notations, $\Omega=(\Sigma_v+D)$, where $D$ is a diagonal matrix with diagonal elements equal to some $\tau_j^2$. 

Now, for proving the required result, we define three different estimators of $\theta$. Below, $\hat\theta$ is the estimator we are considering in this study, $\hat\theta_G$ denotes the usual GLS estimator, and $\hat\theta_F$ is a feasible GLS estimator. 
\begin{align*}
    \hat\theta &= (X\dash\hat\Omega^{-1/2}W\hat\Omega^{-1/2}X)^{-1}(X\dash\hat\Omega^{-1/2}W\hat\Omega^{-1/2}Y) \\
    \hat\theta_G &= (X\dash\Omega^{-1}X)^{-1}(X\dash\Omega^{-1}Y) \\
    \hat\theta_F &= (X\dash\hat\Omega^{-1}X)^{-1}(X\dash\hat\Omega^{-1}Y) 
\end{align*}

In the above, $W$ is the weight matrix as defined in \Cref{subsec:testing} and $\hat\Omega$ is our estimate of the covariance matrix. For convenience, as use $N=nT$ hereafter. Following \citet[Chapter 9]{baltagi11}, we know that $\sqrt N(\hat\theta_G-\theta)$ and $\sqrt N(\hat\theta_F-\theta)$ have the same asymptotic distribution $N(0,\sigma^2Q^{-1})$, where $Q=\lim (X\dash\Omega^{-1}X/N)$ as $N\to\infty$, if $X\dash(\hat\Omega^{-1}-\Omega^{-1})X/N \xrightarrow{P} 0$ and $X\dash(\hat\Omega^{-1}-\Omega^{-1})\varepsilon/N \xrightarrow{P} 0$. Further, a sufficient condition for this to hold is that $\hat\Omega$ is a consistent estimator for $\Omega$ and that $X$ has a satisfactory limiting behavior. 

Let us now assume that the estimate $\hat\tau_j^2$ is consistent for $\tau_j^2$, for all $j$. That would automatically ensure the consistency of $\hat\Omega$ and thereby we can conclude that $\hat\theta_F$ and $\hat\theta_G$ have same asymptotic distribution. Further, note that $X\dash\hat\Omega^{-1/2}W\hat\Omega^{-1/2}X - X\dash\hat\Omega^{-1}X = X\dash\hat\Omega^{-1/2}(W-I)\hat\Omega^{-1/2}X$. Taking any appropriate norm (2-norm, for example) on both sides, we can argue that $\norm{X\dash\hat\Omega^{-1/2}W\hat\Omega^{-1/2}X - X\dash\hat\Omega^{-1}X} \to 0$ as $N\to\infty$, in view of the fact that $\norm{W-I}=2/\log N$, and that $\hat\Omega$ is a consistent estimator for $\Omega$, the population covariance matrix. In a similar fashion, we can show that $\norm{X\dash\hat\Omega^{-1/2}W\hat\Omega^{-1/2}\varepsilon - X\dash\hat\Omega^{-1}\varepsilon} \to 0$ as $N\to\infty$, and thus we can conclude that $\sqrt N(\hat\theta-\theta)$ and $\sqrt N(\hat\theta_F-\theta)$ have the same asymptotic distribution. 

Clearly, all we need to prove is that $\hat\tau_j^2$ is a consistent estimator for $\tau_j^2$ for all $j$. To this end, recall that $\hat\tau_j^2$ is the maximum likelihood estimator (MLE) of $\tau_j^2$ for the problem $\hat\varepsilon_j\sim N(0,(\Sigma_v^{(j)}+\tau_j^2I))$, where $\hat\varepsilon_j$ is the vector of scaled residuals corresponding to the $j$th season and $\Sigma_v^{(j)}$ is the submatrix of $\Sigma_v$ corresponding to the same. It is well-known that MLE is a consistent estimator for such problems. Let $n_j$ be the length of $\varepsilon_j$. Since $T\to\infty$, it is clear that the number of observations per season will also approach infinity, and thus $n_j\to\infty$. Hence, $\hat\tau_j^2$ is consistent for $\tau_j^2$ and that ends our proof for the asymptotic normality of $\hat\theta$. The consistency result follows automatically from the above.
\end{proof}

\bibliography{reference}

\end{document}